\documentclass[12pt,a4paper,oneside]{article}
\usepackage{fullpage}
\usepackage[english]{babel} 
\usepackage[T1]{fontenc} 
\usepackage[utf8]{inputenc} 
\usepackage[babel]{microtype} 

\usepackage{url}
\usepackage{csquotes}
\usepackage{amsmath,amssymb,bm,mathtools}
\usepackage[pdftex]{graphicx}
\graphicspath{{fig/}}
\usepackage{enumitem}
\usepackage{booktabs}
\usepackage{makecell}
\usepackage{siunitx}
\usepackage{caption} 
\usepackage{natbib}
\usepackage{hyperref}

\usepackage{xifthen}
\newcommand{\interank}[1]{%
\ifthenelse{\isempty{#1}}{\textsc{interank}}{\textsc{interank} \emph{#1}}}

\newcommand{\email}[1]{\href{mailto:#1}{\nolinkurl{#1}}}

\newlist{enuminline}{enumerate*}{1}
\setlist[enuminline,1]{label=\itshape\alph*\upshape)}

\title{Can Who-Edits-What Predict Edit Survival?}

\author{
Ali Batuhan Yardım\thanks{
Bilkent University, Ankara, Turkey.
Contact: \email{batuhan.yardim@ug.bilkent.edu.tr}.
This work was done while the author was at EPFL.}
\and
Victor Kristof\thanks{
School of Computer and Communication Sciences, EPFL, Lausanne, Switzerland.
Contact: \email{first.last@epfl.ch}.}
\and
Lucas Maystre\footnotemark[2]
\and
Matthias Grossglauser\footnotemark[2]
}

\begin{document}
\maketitle

\begin{abstract}
As the number of contributors to online peer-production systems grows, it becomes increasingly important to predict whether the edits that users make will eventually be beneficial to the project.
Existing solutions either rely on a user reputation system or consist of a highly specialized predictor that is tailored to a specific peer-production system.
In this work, we explore a different point in the solution space that goes beyond user reputation but does not involve any content-based feature of the edits.
We view each edit as a game between the editor and the component of the project.
We posit that the probability that an edit is accepted is a function of the editor's skill, of the difficulty of editing the component and of a user-component interaction term.
Our model is broadly applicable, as it only requires observing data about \emph{who} makes an edit, \emph{what} the edit affects and whether the edit survives or not.
We apply our model on Wikipedia and the Linux kernel, two examples of large-scale peer-production systems, and we seek to understand whether it can effectively predict edit survival:
in both cases, we provide a positive answer.
Our approach significantly outperforms those based solely on user reputation and bridges the gap with specialized predictors that use content-based features.
It is simple to implement, computationally inexpensive, and in addition it enables us to discover interesting structure in the data.


\end{abstract}

\section{Introduction}
\label{sec:intro}

Over the last two decades, the number and scale of online peer-production systems has become truly massive, driven by better information networks and advances in collaborative software.
At the time of writing, \num{128643} editors contribute regularly to \num{5}+ million articles of the English Wikipedia \citep{wikipedia2017wikipedians} and over \num{15600} developers have authored code for the Linux kernel \citep{corbet2017linux}.
On GitHub, \num{24} million users collaborate on \num{25.3} million active software repositories \citep{github2017octoverse}.

In order to ensure that such projects advance towards their goals, it is necessary to identify whether edits made by users are beneficial.
As the number of users and components of the project grows, this task becomes increasingly challenging.
In response, two types of solutions are proposed.
On the one hand, some advocate the use of \emph{user reputation systems} \citep{resnick2000reputation, adler2007content}.
These systems are general, their predictions are easy to interpret and can be made resistant to manipulations \citep{dealfaro2013content}.
On the other hand, a number of highly specialized methods are proposed to automatically predict the quality of edits in particular peer-production systems \citep{druck2008learning, wikimedia2015artificial}.
These methods can attain excellent predictive performance \citep{heindorf2016vandalism} and usually significantly outperform predictors that are based on user reputation alone \citep{druck2008learning}, but they are tailored to a particular peer-production system, use domain-specific features and rely on models that are difficult to interpret.

In this work, we set out to explore another point in the solution space.
We aim to keep the generality and simplicity of user reputation systems, while reaching the predictive accuracy of highly specialized methods.
We ask the question:
Can one predict the outcome of contributions simply by observing \emph{who edits what} and whether the edits eventually survive?
We address this question by proposing a novel statistical model of edit outcomes.
We formalize the notion of collaborative project as follows.
$N$ users can propose edits on $M$ distinct items (components of the project, such as articles on Wikipedia or a software's modules), and we assume that there is a process for validating edits (either immediately or over time).
We observe triplets $(u, i, q)$ that describe a user $u \in \{1, \ldots, N\}$ editing an item $i \in \{1, \ldots, M\}$ and leading to outcome $q \in \{0, 1\}$;
the outcome $q = 0$ represents a rejected edit, whereas $q = 1$ represents an accepted, beneficial edit.
Given a dataset of such observations, we seek to learn a model of the probability $p_{ui}$ that an edit made by user $u$ on item $i$ is accepted.
This model can then be used to help moderators and project maintainers prioritize their efforts once new edits appear:
For example, edits that are unlikely to survive could be sent out for review immediately.

Our approach borrows from probabilistic models of pairwise comparisons \citep{zermelo1928berechnung, rasch1960probabilistic}.
These models learn a real-valued score for each object (user or item) such that the difference between two objects' scores is predictive of comparison outcomes.
We take a similar perspective and view each edit in a collaborative project as a game between the user who tries to effect change and the item that resists change\footnote{Obviously, items do not really ``resist'' by themselves.
Instead, this notion should be taken as a proxy for the combined action of other users (e.g., project maintainers) who can accept or reject an edit depending, among others, on standards of quality.}.
Similarly to pairwise-comparison models, our approach learns a real-valued score for each user and each item.
In addition, it also learns latent features of users and items that capture interaction effects.

In contrast to quality-prediction methods specialized on a particular peer-production system, our approach is general and can be applied to any system in which users contribute by editing discrete items.
It does not use any explicit content-based features: instead, it simply learns by observing triplets $\{ (u, i, q) \}$.
Furthermore, the resulting model parameters can be interpreted easily.
They enable a principled way of
\begin{enuminline}
\item ranking users by the quality of their contributions,
\item ranking items by the difficulty of editing them and
\item understanding the main dimensions of the interaction between users and items.
\end{enuminline}

We apply our approach on two different peer-production systems.
We start with Wikipedia and consider its Turkish and French editions.
Evaluating the accuracy of predictions on an independent set of edits, we find that our model approaches the performance of the state of the art.
More interestingly, the model parameters reveal important facets of the system.
For example, we characterize articles that are easy or difficult to edit, respectively, and we identify clusters of articles that share common editing patterns.
Next, we turn our attention to the Linux kernel.
In this project, contributors are typically highly skilled professionals, and the edits that they make affect \num{394} different subsystems (kernel components).
In this instance, our model's predictions are \emph{more accurate} than a random forest classifier trained on domain-specific features.
In addition, we give an interesting qualitative description of subsystems based on their difficulty score.

In short, our paper
\begin{enuminline}
\item gives evidence that observing \emph{who edits what} can yield valuable insights into peer-production systems and
\item proposes a statistically grounded and computationally inexpensive method to do so.
\end{enuminline}
The analysis of two peer-production systems with very distinct characteristics demonstrates the generality of the approach.

\paragraph{Organization of the Paper}
We start by reviewing related literature in Section~\ref{sec:relwork}.
In Section~\ref{sec:models}, we describe our statistical model of edit outcomes and briefly discuss how to efficiently learn a model from data.
In Sections~\ref{sec:wikipedia} and~\ref{sec:linux}, we investigate our approach in the context of Wikipedia and of the Linux kernel, respectively.
Finally, we conclude in Section~\ref{sec:conclusion}.

\section{Related Work}
\label{sec:relwork}

With the growing size and impact of online peer-production systems, the task of assessing contribution quality has been extensively studied.
We review various approaches to the problem of quantifying and predicting the quality of user contributions and contrast them to our approach.

\paragraph{User Reputation Systems}
Reputation systems have been a long-standing topic of interest in relation to peer-production systems and, more generally, in relation to online services \citep{resnick2000reputation}.
\citet{adler2007content} propose a point-based reputation system for Wikipedia and show that reputation scores are predictive of the future quality of editing.
As almost all edits to Wikipedia are immediately accepted, the authors define an \emph{implicit} notion of edit quality by measuring how much of the introduced changes is retained in future edits.
The ideas underpinning the computation of implicit edit quality are extended and refined in subsequent papers \citep{adler2008measuring, dealfaro2013content}.
This line of work leads to the development of WikiTrust \citep{dealfaro2011reputation}, a browser add-on that highlights low-reputation texts in Wikipedia articles.
When applying our methods to Wikipedia, we follow the same idea of measuring quality implicitly through the state of the article at subsequent revisions.
We also demonstrate that by automatically learning properties of the \emph{item} that a user edits (in addition to learning properties of the user, such as a reputation score) we can substantially improve predictions of edit quality.
This was also noted recently by \citet{tabibian2017distilling} in a setting similar to ours, but using a temporal point process framework.

\paragraph{Specialized Classifiers}
Several authors propose quality-prediction methods tailored to a specific peer-production system.
Typically, these methods consist of a machine-learned classifier trained on a large number of content-based and system-based features of the users, the items and the edits themselves.
\citet{druck2008learning} fit a maximum entropy classifier for estimating the lifespan of a given Wikipedia edit, using a definition of edit longevity similar to that of \citet{adler2007content}.
They consider features based on the edit's content (such as: number of words added / deleted, type of change, capitalization and punctuation, etc.) as well as features based on the user, the time of the edit and the article.
Their model significantly outperforms a baseline that only uses features of the user.
Other methods use support vector machines \citep{bronner2012user}, random forests \citep{bronner2012user, javanmardi2011vandalism} or binary logistic regression \citep{potthast2008automatic}, with varying levels of success.
In some cases, content-based features are refined using natural-language processing, leading to substantial performance improvements.
However, these improvements are made to the detriment of general applicability.
For example, competitive natural language processing tools have yet to be developed for the Turkish language (we investigate the Turkish Wikipedia in Section~\ref{sec:wikipedia}).
In contrast to these methods, our approach is general and broadly applicable.
Furthermore, the use of black-box classifiers can hinder the interpretability of predictions, whereas we propose a statistical model whose parameters are straightforward to interpret.

\paragraph{Truth Inference}
In crowdsourcing, a problem related to ours consists of \emph{jointly} estimating
\begin{enuminline}
\item model parameters (such as user skills or item difficulties) that are predictive of contribution quality, and
\item the quality of each contribution,
\end{enuminline}
without ground truth \citep{dawid1979maximum}.
Our problem is therefore easier, as we assume access to ground-truth information about the outcome (quality) of past edits.
Nevertheless, some methods developed in the crowdsourcing context \citep{whitehill2009whose, welinder2010multidimensional, zhou2012learning} provide models that can be applied to our setting as well.
In Sections~\ref{sec:wikipedia} and~\ref{sec:linux}, we compare our models to GLAD \citep{whitehill2009whose}.


\paragraph{Pairwise Comparison Models}
Our approach draws inspiration from probabilistic models of pairwise comparisons.
These have been studied extensively over the last century in the context of psychometrics \citep{thurstone1927law, bradley1952rank}, item response theory \citep{rasch1960probabilistic}, chess rankings \citep{zermelo1928berechnung, elo1978rating}, and more.
The main paradigm posits that every object $i$ has a latent \emph{strength} (skill or difficulty) parameter $\theta_i$, and that the probability $p_{ij}$ of observing object $i$ ``winning'' over object $j$ increases with the distance $\theta_i - \theta_j$.
Conceptually, our model is closest to that of \citet{rasch1960probabilistic}.

\paragraph{Collaborative Filtering}
Our method also borrows from collaborative filtering techniques popular in the recommender systems community.
In particular, some parts of our model are remindful of matrix-factorization techniques \citep{koren2009matrix}.
These techniques automatically learn low-dimensional embeddings of users and items based on ratings, with the purpose of producing better recommendations.
Our work shows that these ideas can also be helpful in addressing the problem of predicting outcomes of edits in peer-production systems.
Like collaborative-filtering methods, our approach is exposed to the \emph{cold-start} problem:
with no (or few) observations about a given user or item, the predictions are notably less accurate.
In practice, this problem can be addressed, e.g., by using additional features of users and / or items \citep{schein2002methods, lam2008addressing} or by clustering users \citep{levi2012finding}.

\section{Statistical Models}
\label{sec:models}

In this section, we describe and explain two variants of a statistical model of edit outcomes based on \emph{who} edits \emph{what}.
In other words, we develop models that are predictive of the outcome $q \in \{0, 1\}$ of a contribution of user $u$ on item $i$.
To this end, we represent the probability $p_{ui}$ that an edit made by user $u$ on item $i$ is successful.
In collaborative projects of interest, most users typically interact with only a small number of items.
In order to deal with the sparsity of interactions, we postulate that the probabilities $\{ p_{ui} \}$ lie on a low-dimensional manifold and propose two model variants of increasing complexity.
In both cases, the parameters of the model have intuitive effects and can be interpreted easily.

\paragraph{Basic Variant}
The first variant of our model is directly inspired by the Rasch model \citep{rasch1960probabilistic}.
The probability that an edit is accepted is defined as
\begin{align}
\label{eq:basicmodel}
p_{ui} = \frac{1}{1 + \exp[-(s_u - d_i + b)]},
\end{align}
where $s_u\in\mathbf{R}$ is the \emph{skill} of user $u$, $d_i\in\mathbf{R}$ is the \emph{difficulty} of item $i$, and $b \in \mathbf{R}$ is a global parameter that encodes the overall skew of the distribution of outcomes.
We call this model variant \interank{basic}.
Intuitively, the model predicts the outcome of a ``game'' between an item with inertia and a user who would like to effect change.
The \emph{skill} quantifies the ability of the user to enforce a contribution, whereas the \emph{difficulty} quantifies how ``resistant'' to contributions the particular item is.

Similarly to reputation systems \citep{adler2007content}, \interank{basic} learns a score for each user; this score is predictive of edit quality.
However, unlike these systems, our model also takes into account that some items might be more challenging to edit than others.
For example, on Wikipedia, we can expect high-traffic, controversial articles to be more difficult to edit than less popular articles.
As with user skills, the article difficulty can be inferred \emph{automatically} from observed outcomes.

\paragraph{Full Variant}
Although the \emph{basic} variant is conceptually attractive, it might prove to be too simplistic in some instances.
In particular, the \emph{basic} variant implies that if user $u$ is more skilled than user $v$, then $p_{ui} > p_{vi}$ for \emph{all} items $i$.
In many peer-production systems, users tend to have their own specializations and interests, and each item in the project might require a particular mix of skills.
For example, with the Linux kernel, an engineer specialized in file systems might be successful in editing a certain subset of software components, but might be less proficient in contributing to, say, network drivers, whereas the situation might be exactly the opposite for another engineer.
In order to capture the multidimensional interaction between users and items, we add a bilinear term to the probability model~\eqref{eq:basicmodel}.
Letting $\bm{x}_u, \bm{y}_i \in \mathbf{R}^D$ for some dimensionality $D \in \mathbf{N}_{>0}$, we define
\begin{align}
\label{eq:fullmodel}
p_{ui} = \frac{1}{1 + \exp[-(s_u - d_i + \bm{x}_u^\top \bm{y}_i + b)]}.
\end{align}
We call the corresponding model variant \interank{full}.
The vectors $\bm{x}_u$ and $\bm{y}_i$ can be thought of as embedding users and items as points in a latent $D$-dimensional space.
Informally, $p_{ui}$ increases if the two points representing a user and an item are close to each other, and it decreases if they are far from each other (e.g., if the vectors have opposite signs).
If we slightly oversimplify, the parameter $\bm{y}_i$ can be interpreted as describing the set of skills needed to successfully edit item $i$, whereas $\bm{x}_u$ describes the set of skills displayed by user~$u$.

The bilinear term is reminiscent of matrix-factorization approaches in recommender systems \citep{koren2009matrix};
indeed, this variant can be seen as a \emph{collaborative-filtering} method.
In true collaborative-filtering fashion, our model is able to learn the latent feature vectors $\{ \bm{x}_i \}$ and $\{ \bm{y}_i \}$ \emph{jointly}, by taking into consideration all edits and without any additional content-based features.

Finally, note that the skill and difficulty parameters are retained in this variant and can still be used to explain first-order effects.
The bilinear term explains only the additional effect due to the user-item interaction.

\subsection{Learning the Model}
\label{sec:learning}

From~\eqref{eq:basicmodel} and~\eqref{eq:fullmodel}, it should be clear that our probabilistic model assumes no data other than the identity of the user and that of the item.
This makes it generally applicable to any peer-production system in which users contribute to discrete items.

Given a dataset of $K$ independent observations $\mathcal{D} = \{ (u_k, i_k, q_k) \mid k = 1, \ldots, K \}$, we infer the parameters of the model by maximizing their likelihood under $\mathcal{D}$.
That is, collecting all model parameters into a single vector $\bm{\theta}$, we seek to minimize the negative log-likelihood
\begin{align}
\label{eq:nll}
- \ell (\bm{\theta} ; \mathcal{D}) = \sum_{(u,i,q) \in \mathcal{D}} \left[ -q \log p_{ui} - (1 - q) \log (1 - p_{ui}) \right],
\end{align}
where $p_{ui}$ depends on $\bm{\theta}$.
In the \emph{basic} variant, the negative log-likelihood is convex, and we can easily find a global maximum by using standard methods from convex optimization.
In the \emph{full} variant, the bilinear term breaks the convexity of the objective function, and we can no longer guarantee that we will find parameters that are global minimizers.
In practice, we do not observe any convergence issues but reliably find good model parameters on all datasets.

Note that~\eqref{eq:nll} easily generalizes from binary outcomes ($q \in \{0, 1\}$) to continuous-valued outcomes ($q \in [0, 1]$).
Continuous values can be used to represent the \emph{fraction} of the edit that is successful.

\paragraph{Implementation}
We implement the models in Python by using the TensorFlow library \citep{abadi2016tensorflow}.
Our code is publicly available online at \url{https://github.com/lca4/interank}.
In order to avoid overfitting the model to the training data, we add a small amount of $\ell_2$ regularization to the negative log-likelihood.
We minimize the negative log-likelihood by using stochastic gradient descent \citep{bishop2006pattern} with small batches of data.
For \interank{full}, we set the number of latent dimensions to $D = 20$ by cross-validation.

\paragraph{Running Time}
Our largest experiment consists of learning the parameters of \interank{full} on the entire history of the French Wikipedia (c.f. Section~\ref{sec:wikipedia}), consisting of over \num{65} million edits by \num{5} million users on \num{2} million items.
In this case, our TensorFlow implementation takes approximately \num{2} hours to converge on a single machine.
In most other experiments, our implementation takes only a few minutes to converge.
This demonstrates that our model effortlessly scales, even to the largest peer-production systems.

\subsection{Applicability}
Our approach models the difficulty of effecting change through the affected item's identity.
As such, it applies particularly well to peer-production systems where users \emph{cooperate} to improve the project, i.e., where each edit is judged independently against an item's (latent) quality standards.
This model is appropriate for a wide variety of projects, ranging from online knowledge bases (such as Wikipedia, c.f. Section~\ref{sec:wikipedia}) to open source software (such as the Linux kernel project, c.f. Section~\ref{sec:linux}).
In some peer-production systems, however, the contributions of different users \emph{compete} against each other, such as multiple answers to a single question on a Q\&A platform.
In these cases, our model can still be applied, but fails to capture the fact that edit outcomes are interdependent.

\begin{table*}
  \caption{Summary statistics of Wikipedia datasets after preprocessing.}
  \label{tab:wikidata}
  \centering
  \begin{footnotesize}
  \begin{tabular}{lrrrrrrr}
    \toprule
    Edition &  \# users $N$ & \# articles $M$ &       \# edits & First edit &  Last edit & $q < 0.2$ & $q > 0.8$ \\
    \midrule
    French  & \num{5460745} &   \num{1932810} & \num{65430838} & 2001-08-04 & 2017-09-02 &            \num{6.4} \% &           \num{72.2} \% \\
    Turkish & \num{1360076} &    \num{310991} &  \num{8768258} & 2002-12-05 & 2017-10-01 &           \num{11.6} \% &           \num{60.5} \% \\
    \bottomrule
  \end{tabular}
  \end{footnotesize}
\end{table*}

\section{Wikipedia}
\label{sec:wikipedia}

Wikipedia is a popular free online encyclopedia and arguably one of the most successful peer-production systems.
In this section, we apply our models to the French and Turkish editions of Wikipedia.

\subsection{Background \& Datasets}

The French Wikipedia is one of the largest Wikipedia editions.
At the time of writing, it ranks in third position both in terms of number of edits and number of users\footnote{%
We chose the French edition over the English one because our computing infrastructure could not support the $\approx15$ TB needed to store the entire history of the English Wikipedia.
The French edition contains roughly $5\times$ fewer edits.
}.
In order to obtain a complementary perspective, we also study the Turkish Wikipedia, which is roughly an order of magnitude smaller.
Interestingly, both the French and the Turkish editions score very highly on Wikipedia's \emph{depth} scale, a measure of collaborative quality \citep{wikimedia2017depth}.

The Wikimedia Foundation releases periodically and publicly a database dump containing the successive revisions to all articles\footnote{%
See: \url{https://dumps.wikimedia.org/}.}.
In this paper, we use a dump that contains data starting from the beginning of the edition up to the fall of 2017.

\subsubsection{Computation of Edit Quality}

On Wikipedia, any user's edit is immediately incorporated into the encyclopedia\footnote{Except for a small minority of protected articles.}.
Therefore, in order to obtain information about the quality of an edit, we have to consider the implicit signal given by subsequent edits to the same article.
If the changes introduced by the edit are preserved, it signals that the edit was beneficial, whereas if the changes are reverted, the edit likely had a negative effect.
A formalization of this idea is given by \citet{adler2007content} and \citet{druck2008learning};
see also \citet{dealfaro2013content} for a concise explanation.
In this paper, we essentially follow their approach.

Consider a particular article and denote by $v_k$ its $k$-th revision (i.e., the state of the article after the $k$-th edit).
Let $d(u, v)$ be the Levenshtein distance between two revisions \citep{kruskal1983overview}.
We define the \emph{quality} of edit $k$ from the perspective of the article's state after $\ell \ge 1$ subsequent edits as
\begin{align*}
q_{k \mid \ell} = \frac{1}{2} + \frac{d(v_{k-1}, v_{k + \ell}) - d(v_k, v_{k + \ell})}{2 d(v_{k-1}, v_k)}.
\end{align*}
By properties of distances, $q_{k \mid \ell} \in [0, 1]$.
Intuitively, the quantity $q_{k \mid \ell}$ captures the proportion of work done in edit $k$ that remains in revision $k + \ell$.
It can be understood as a \emph{soft} measure of whether edit $k$ has been reverted or not.
We compute the unconditional quality of the edit by averaging over multiple future revisions:
\begin{align}
\label{eq:wikiqual}
q_k = \frac{1}{L} \sum_{\ell = 1}^L q_{k \mid \ell},
\end{align}
where $L$ is the minimum between the number of subsequent revisions of the article and $10$ (we empirically found that \num{10} revisions is enough to accurately assess the quality of an edit).
Note that even though $q_k$ is no longer binary, our models naturally extend to continuous-valued $q_k \in [0,1]$ (c.f. Section~\ref{sec:learning}).

In practice, we observe that edit quality is bimodal and asymmetric.
Most edits have a quality close to either \num{0} or \num{1} and a majority of edits are of high quality.
The two rightmost columns of Table~\ref{tab:wikidata} quantify this for the French and Turkish editions.

\subsubsection{Dataset Preprocessing}

We consider all edits to the pages in the main namespace (i.e., articles), including those from anonymous contributors identified by their IP address\footnote{%
Note, however, that a large majority of edits are made by registered users (\num{82.7} \% and \num{76.6} \% for the French and Turkish editions, respectively).}.
Sequences of consecutive edits to an article by the same user are collapsed into a single edit in order to remove bias in the computation of edit quality \citep{adler2007content}.
To evaluate methods in a realistic setting, we split the data into a training set containing the first \num{90} \% of edits, and we report results on an independent validation set containing the remaining \num{10} \%.
Note that the quality is computed based on subsequent revisions of an article:
In order to guarantee that the two sets are truly independent, we make sure that we never use any revisions from the validation set to compute the quality of edits in the training set.
A short summary of the data statistics after preprocessing is provided in Table~\ref{tab:wikidata}.

\subsection{Evaluation}

In order to facilitate the comparison of our method with competing approaches, we evaluate the performance on a binary classification task consisting of predicting whether an edit is of poor quality.
To this end, we assign binary labels to all edits in the validation set: the label \emph{bad} is assigned to every edit with $q < 0.5$, and the label \emph{good} is assigned to all edits with $q \ge 0.5$.
The predictions of the classifier might help Wikipedia administrators to identify edits of low quality;
these edits might then be sent to domain experts for review.

As discussed in Section~\ref{sec:models}, we consider two versions of our model.
The first one, \textsc{interank} \emph{basic}, simply learns scalar user skills and article difficulties.
The second one, \textsc{interank} \emph{full}, additionally includes a latent embedding of dimension $D = 20$ for each user and article.

\subsubsection{Competing Approaches}
\label{sec:wikicompeting}
To set our results in context, we compare them to those obtained with four different baselines.

\paragraph{Average}
The first approach always outputs the marginal probability of a bad edit in the training set, i.e.,
\begin{align*}
p = \frac{\text{\# bad edits in training set}}{\text{\# edits in training set}}
\end{align*}
This is a trivial baseline, and it gives an idea of what results we should expect to achieve without any additional information on the user, article or edit.

\paragraph{User-Only}
The second approach models the outcome of an edit using only the user's identity.
In short, the predictor learns skills $\{s_u \mid u = 1, \ldots, N\}$ and a global offset $b$ such that, for each user $u$, the probability
\begin{align*}
p_u = \frac{1}{1 + \exp[- (s_u + b)]}
\end{align*}
maximizes the likelihood of that user's edits in the training set.
This baseline predictor is representative of user reputation systems such as that of \citet{adler2007content}.

\paragraph{GLAD}
In the context of crowdsourcing, \citet{whitehill2009whose} propose the GLAD model that postulates that
\begin{align*}
p_{ui} = \frac{1}{1 + \exp(- s_u / d_i)},
\end{align*}
where $s_u \in \mathbf{R}$ and $d_i \in \mathbf{R}_{>0}$.
This reflects a different assumption on the interplay between user skill and item difficulty: under their model, an item with a large difficulty value makes every user's skill more ``diffuse''.
In order to make the comparison fair, we add a global offset parameter $b$ to the model (similarly to \textsc{interank} and the user-only baseline).

\paragraph{ORES reverted}
The fourth approach is a state-of-the-art classifier developed by researchers at the Wikimedia Foundation as part of Wikipedia's Objective Revision Evaluation Service \citep{wikimedia2015artificial}.
We use the two classification models specifically developed for the French and Turkish editions.
Both models use over \num{80} content-based and system-based features extracted from the user, the article and the edit to predict whether the edit will be reverted, a target which essentially matches our operational definition of \emph{bad} edit.
Features include the number of vulgar words introduced by the edit, the length of the article and of the edit, etc.
This predictor is representative of specialized, domain-specific approaches to modeling edit quality.

\subsubsection{Results}

\begin{figure*}
\centering
\includegraphics[scale=1.2]{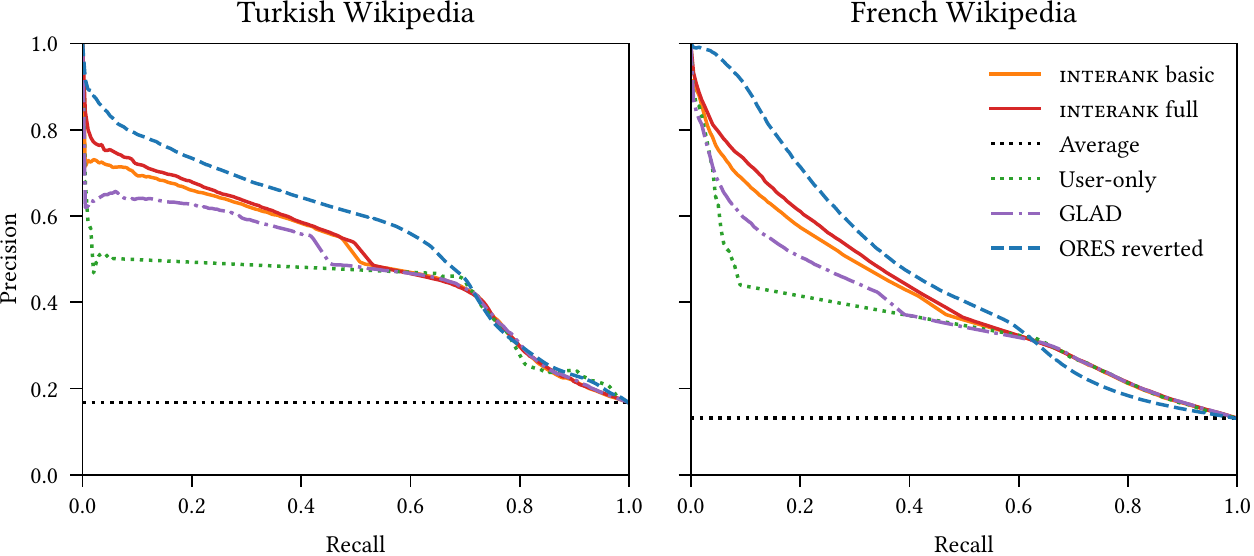}
\includegraphics[scale=1.2]{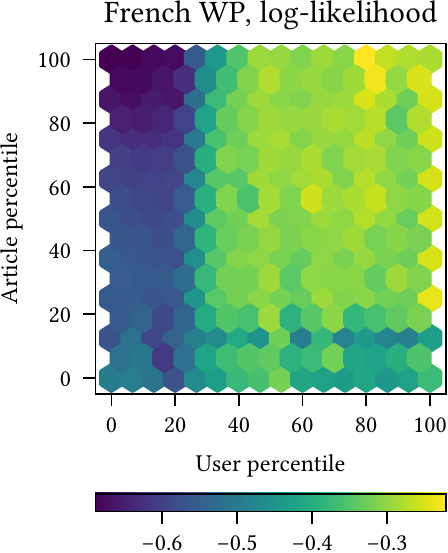}
\caption{Precision-recall curves on the \emph{bad edit} classification task for the Turkish and French editions of Wikipedia (top).
Average log-likelihood as a function of the number of observations of the user and item in the training set (bottom).}
\label{fig:wikipr}
\end{figure*}

Table~\ref{tab:wikiperf} presents the average log-likelihood and the area under the precision-recall curve (AUPRC) for each method.
\textsc{interank} \emph{full} has the highest average log-likelihood of all models, meaning that its predictive probabilities are well calibrated with respect to the validation data.

\begin{table}
  \caption{Predictive performance on the \emph{bad edit} classification task for the French and Turkish editions of Wikipedia.
The best performance is highlighted in bold.}
  \label{tab:wikiperf}
  \centering
  \begin{tabular}{llrr}
    \toprule
    Edition        & Model                          & Avg. log-likelihood   & AUPRC \\
    \midrule
    French         & \textsc{interank} \emph{basic} & \num{-0.339}          & \num{0.399}  \\
                   & \textsc{interank} \emph{full}  & \textbf{\num{-0.336}} & \num{0.413}  \\
    \addlinespace
                   & Average                        & \num{-0.389}          & \num{0.131}  \\
                   & User-only                      & \num{-0.346}          & \num{0.313}  \\
                   & GLAD                           & \num{-0.344}          & \num{0.369} \\
                   & ORES reverted                  & \num{-0.469}          & \textbf{\num{0.453}} \\
    \midrule
    Turkish        & \textsc{interank} \emph{basic} & \num{-0.380}          & \num{0.494}  \\
                   & \textsc{interank} \emph{full}  & \textbf{\num{-0.379}} & \num{0.503}  \\
    \addlinespace
                   & Average                        & \num{-0.461}          & \num{0.168}  \\
                   & User-only                      & \num{-0.390}          & \num{0.410}  \\
                   & GLAD                           & \num{-0.387}          & \num{0.471} \\
                   & ORES reverted                  & \num{-0.392}          & \textbf{\num{0.552}} \\
    \bottomrule
  \end{tabular}
\end{table}

Figure~\ref{fig:wikipr} (top) presents the precision-recall curves for all methods.
The analysis is qualitatively similar for both Wikipedia editions.
All non-trivial predictors perform similarly in the high-recall regime, but present significant differences in the high-precision regime, on which we will focus.
The ORES predictor performs the best.
\textsc{interank} comes second, reasonably close behind ORES, and the \emph{full} variant has a small edge over the \emph{basic} variant.
GLAD is next, and the user-only baseline is far behind.
This shows that
\begin{enuminline}
\item incorporating information about the article being edited is crucial for achieving a good performance on a large portion of the precision-recall trade-off, and
\item modeling the outcome probability by using the \emph{difference} between skill and difficulty (\textsc{interank}) is better than by using the \emph{ratio} (GLAD).
\end{enuminline}

We also note that in the validation set, approximately \num{20} \% (\num{15} \%) of edits are made by users (respectively, on articles) that are never encountered in the training set (the numbers are similar in both editions).
In these cases, \textsc{interank} reverts to average predictions, whereas content-based methods can take advantage of other features of the edit to make an informed prediction.
In order to explore this \emph{cold-start} effect in more detail, we group users and articles into bins based on the number of times they appear in the training set, and we compute the average log-likelihood of validation examples separately for each bin.
Figure~\ref{fig:wikipr} (bottom) presents the results for the French edition;
the results for the Turkish edition are similar.
Clearly, predictions for users and articles present in the training set are significantly better.
In a practical deployment, several methods can help to address this issue \citep{schein2002methods, lam2008addressing, levi2012finding}.
A thorough investigation of ways to mitigate the cold-start problem is beyond the scope of this paper.

In summary, we observe that our model, which incorporates the articles' identity, is able to bridge the gap between user-only prediction approach and a specialized predictor (ORES reverted).
Furthermore, modeling the interaction between user and article (\textsc{interank} \emph{full}) is beneficial and helps further improve predictions, particularly in the high-precision regime.

\subsection{Interpretation of Model Parameters}

The parameters of \textsc{interank} models, in addition to being predictive of edit outcomes, are also very interpretable.
In the following, we demonstrate how they can surface interesting characteristics of the peer-production system.

\subsubsection{Controversial Articles}
Intuitively, we expect an article $i$ whose difficulty parameter $d_i$ is large to deal with topics that are potentially controversial.
We focus on the French Wikipedia and explore a list of the ten most controversial articles given by \citet{yasseri2014most}.
In this 2014 study, the authors identify controversial articles by using an ad-hoc methodology.
Table~\ref{tab:wikicontrov} presents, for each article identified by \citeauthor{yasseri2014most}, the percentile of the corresponding difficulty parameter $d_i$ learned by \textsc{interank} \emph{full}.
We analyze these articles approximately four years later, but the model still identifies them as some of the most difficult ones.
Interestingly, the article on Sigmund Freud, which has the lowest difficulty parameter of the list, has become a \emph{featured} article since \citeauthor{yasseri2014most}'s analysis---a distinction awarded only to the most well-written and neutral articles.

\begin{table}
  \caption{The ten most controversial articles on the French Wikipedia according to \citet{yasseri2014most}.
For each article $i$, we indicate the percentile of its corresponding parameter $d_i$.}
  \label{tab:wikicontrov}
  \centering
  \begin{tabular}{rlr}
    \toprule
    Rank & Title                       & Percentile of $d_i$ \\
    \midrule
       1 & Ségolène Royal              & \num{99.840} \% \\
       2 & Unidentified flying object  & \num{99.229} \% \\
       3 & Jehovah's Witnesses         & \num{99.709} \% \\
       4 & Jesus                       & \num{99.953} \% \\
       5 & Sigmund Freud               & \num{97.841} \% \\
       6 & September 11 attacks        & \num{99.681} \% \\
       7 & Muhammad al-Durrah incident & \num{99.806} \% \\
       8 & Islamophobia                & \num{99.787} \% \\
       9 & God in Christianity         & \num{99.712} \% \\
      10 & Nuclear power debate        & \num{99.304} \% \\
    \addlinespace
         & \emph{median}               & \num{99.710} \% \\
    \bottomrule
  \end{tabular}
\end{table}

\subsubsection{Latent Factors}

\begin{table*}
  \caption{A selection of articles of the Turkish Wikipedia among the top-\num{20} highest and lowest coordinates along the first principal axis of the matrix $\bm{Y}$.}
  \label{tab:wikitrlatent}
  \centering
  \begin{tabular}{lp{5in}}
    \toprule
    Direction & Titles \\
    \midrule
    Lowest  &
        Harry Potter's magic list,
        List of programs broadcasted by Star TV,
        Bursaspor 2011-12 season,
        Kral Pop TV Top 20,
        Death Eater,
        Heroes (TV series),
        List of programs broadcasted by TV8,
        Karadayı,
        Show TV,
        List of episodes of Kurtlar Vadisi Pusu. \\
    Highest &
        Seven Wonders of the World,
        Thomas Edison,
        Cell,
        Mustafa Kemal Atatürk,
        Albert Einstein,
        Democracy,
        Isaac Newton,
        Mehmed the Conqueror,
        Leonardo da Vinci,
        Louis Pasteur. \\
    \bottomrule
  \end{tabular}
\end{table*}

Next, we turn our attention to the parameters $\{ \bm{y}_i \}$.
These parameters can be thought of as an embedding of the articles in a latent space of dimension $D = 20$.
As we learn a model that maximizes the likelihood of edit outcomes, we expect these embeddings to capture latent article features that explain edit outcomes.
In order to extract the one or two directions that explain most of the variability in this latent space, we apply principal component analysis \citep{bishop2006pattern} to the matrix $\bm{Y} = [\bm{y}_i]$.

In Table~\ref{tab:wikitrlatent}, we consider the Turkish Wikipedia and list a subset of the \num{20} articles with the highest and lowest coordinates along the first principal axis of $\bm{Y}$.
We observe that this axis seems to distinguish articles about popular culture from those about ``high culture'' or timeless topics.
This discovery supports the hypothesis that users have a propensity to successfully edit \emph{either} popular culture \emph{or} high-culture articles on Wikipedia, but not \emph{both}.

\begin{figure}
\centering
\includegraphics[width=0.6\linewidth]{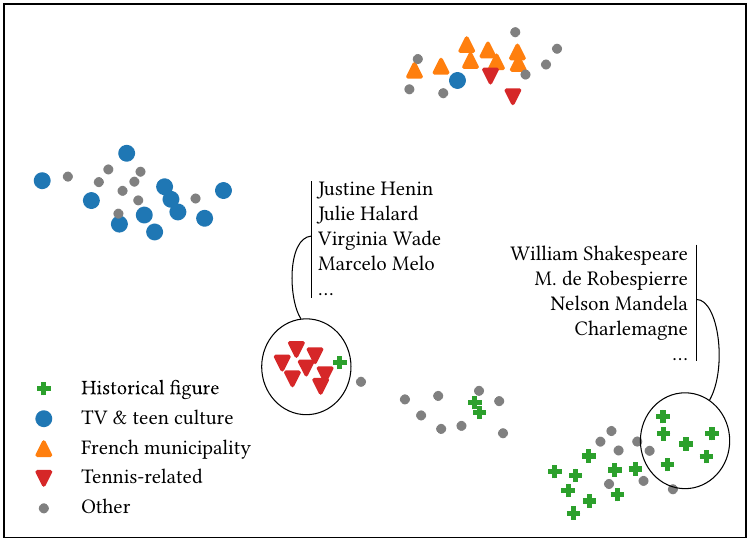}
\caption{$t$-SNE visualization of 80 articles of the French Wikipedia with highest and lowest coordinates along the first and second principal axes of the matrix $\bm{Y}$.}
\label{fig:wikifrlatent}
\end{figure}

Finally, we consider the French Wikipedia.
Once again, we apply principal component analysis to the matrix $\bm{Y}$ and keep the first two dimensions.
We select the \num{20} articles with the highest and lowest coordinates along the first two principal axes\footnote{%
Interestingly, the first dimension has a very similar interpretation to that obtained on the Turkish edition: it can also be understood as separating popular culture from high culture.}.
A two-dimensional $t$-SNE plot \citep{vandermaaten2008visualizing} of the 80 articles selected using PCA is displayed in Figure~\ref{fig:wikifrlatent}.
The plot enables identifying meaningful clusters of related articles, such as articles about tennis players, French municipalities, historical figures, and TV or teen culture.
These articles are representative of the latent dimensions that separate editors the most:
a user skilled in editing pages about ancient Greek mathematicians might be less skilled in editing pages about \emph{anime}, and vice versa.

\section{Linux Kernel}
\label{sec:linux}

In this section, we apply the \interank{} model to the Linux kernel project, a well-known open-source software project.
In contrast to Wikipedia, most contributors to the Linux kernel are highly skilled professionals who dedicate a significant portion of their time and efforts to the project.

\subsection{Background \& Dataset}

The Linux kernel has fundamental impact on technology as a whole.
In fact, the Linux operating system runs 90 \% of the cloud workload and 82 \% of the smartphones \citep{corbet2017linux}.
To collectively improve the source code, developers submit bug fixes or new features in the form of a \emph{patch} to collaborative repositories.
Review and integration time depend on the project's structure, ranging from a few hours or days for Apache Server \citep{rigby2008open} to a couple of months for the Linux kernel \citep{jiang2013will}.
In particular for the Linux kernel, developers submit patches to subsystem mailing lists, where they undergo several rounds of reviews.
After suggestions are implemented and if the code is approved, the patch can be committed to the subsystem maintainer's software repository.
Integration conflicts are spotted at this stage by other developers monitoring the maintainer's repository and any issues must be fixed by the submitter.
If the maintainer is satisfied with the patch, she commits it to Linus Torvalds' repository, who decides to include it or not with the next Linux release.

\subsubsection{Dataset Preprocessing}

We use a dataset collected by \citet{jiang2013will} which spans Linux development activity between 2005 and 2012\footnote{%
The dataset is publicly available online at \url{http://mcis.polymtl.ca/publications/2013/linux_patch_final.arff.gz}.}.
It consists of \num{670 533} patches described using \num{62} features derived from e-mails, commits to software repositories, the developers' activity and the content of the patches themselves.
\citeauthor{jiang2013will} scraped patches from the various mailing lists and matched them with commits in the main repository.
In total, they managed to trace back 75 \% of the commits that appear in Linus Torvalds' repository to a patch submitted to a mailing list.
A patch is labeled as \emph{accepted} ($q = 1$) if it eventually appears in a release of the Linux kernel, and \emph{rejected} ($q = 0$) otherwise.
We remove data points with empty subsystem and developer names, as well as all subsystems with no accepted patches.
Finally, we chronologically order the patches according to their mailing list submission time.

After preprocessing, the dataset contains $K= \num{619419}$ patches proposed by $ N = \num{9672} $ developers on $M = \num{394}$ subsystems.
34.12 \% of these patches are accepted.
We then split the data into training set containing the first 80 \% of patches and a validation set containing the remaining 20 \%.

\subsubsection{Subsystem-Developer Correlation}

\begin{figure}
  \centering
  \includegraphics[width=0.6\linewidth]{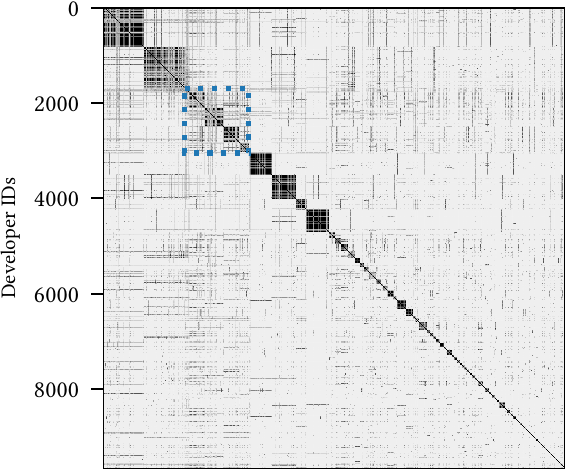}
  \caption{Correlation matrix $ \bm{R} $ between developers ordered according to the subsystem they contribute to the most. The blocks on the diagonal correspond to subsystems.
  Core subsystems form a strong cluster (blue square).}
  \label{fig:linux_correlation}
\end{figure}

Given the highly complex nature of the project, one could believe that developers tend to specialize in few, independent subsystems.
Let $X_u = \{ X_{ui} \}_{i=1}^M$ be the collection of binary variables $ X_{ui} $ indicating whether developer $u$ has an accepted patch in subsystem $i$.
We compute the sample Pearson correlation coefficient $r_{uv} = \rho(X_u, X_v)$ between $X_u$ and $X_v$.
We show in Figure \ref{fig:linux_correlation} the correlation matrix $ \bm{R} = [r_{uv}] $ between developers patching subsystems.
Row $\bm{r}_u$ corresponds to developer $u$, and we order all rows according to the subsystem each developer $u$ contribute to the most.
We order the subsystems in decreasing order by the number of submitted patches, such that larger subsystems appear at the top of the matrix $\bm{R}$.
Hence, the blocks on the diagonal roughly correspond to subsystems and their size represents the number of developers involved with the subsystem.
As shown by the blocks, developers tend to specialize into one subsystem.
However, as the numerous non-zero off-diagonal entries reveal, they still tend to contribute substantially to other subsystems.
Finally, as highlighted by the dotted, blue square, subsystems number three to six on the diagonal form a cluster.
In fact, these four subsystems (\texttt{include/linux}, \texttt{arch/x86}, \texttt{kernel} and \texttt{mm}) are core subsystems of the Linux kernel.

\subsection{Evaluation}

We consider the task of predicting whether a patch will be integrated into a release of the kernel.
Similarly to Section \ref{sec:wikipedia}, we use \interank{basic} and \interank{full} with $D = 20$ latent dimensions to learn the developers' skills, the subsystems' difficulty, and the interaction between them.

\subsubsection{Competing Approaches}
Three baselines that we consider---\emph{average}, \emph{user-only} and \emph{GLAD}---are identical to those described in Section~\ref{sec:wikicompeting}.
In addition, we also compare our model to a random forest classifier trained on domain-specific features similar to the one used by \citet{jiang2013will}.
In total, this classifier has access to 21 features for each patch.
Features include information about the developer's experience up to the time of submission (e.g., number of accepted commits, number of patches sent), the e-mail thread (e.g., number of developers in copy of the e-mail, size of e-mail, number of e-mails in thread until the patch) and the patch itself (e.g., number of lines changed, number of files changed).
We optimize the hyperparameters of the random forest using a grid-search.
As the model has access to domain-specific features about each edit, it is representative of the class of specialized methods tailored to the Linux kernel peer-production system.

\subsubsection{Results}

Table \ref{tab:linux_results} displays the average log-likelihood and area under the precision-recall curve (AUPRC).
\interank{full} performs best in terms of both metrics.
In terms of AUPRC, it outperforms the random forest classifier by 4.4 \%, GLAD by 5 \%, and the \emph{user-only} baseline by 7.3 \%.

\begin{table}
  \caption{Predictive performance on the \emph{accepted patch} classification task for the Linux kernel.
  The best performance is highlighted in bold.}
  \label{tab:linux_results}
  \centering
  \begin{tabular}{lrr}
    \toprule
    Model            & Avg. log-likelihood & AUPRC \\
    \midrule
    \interank{basic} & -0.589              & 0.525 \\
    \interank{full}  & \textbf{-0.588}     & \textbf{0.527} \\
    \addlinespace
    Average          & -0.640              & 0.338 \\
    User-only        & -0.601              & 0.491 \\
    GLAD             & -0.598              & 0.502 \\
    Random forest    & -0.599              & 0.505 \\
    \bottomrule
  \end{tabular}
\end{table}

\begin{figure}
  \centering
  \includegraphics[width=0.6\linewidth]{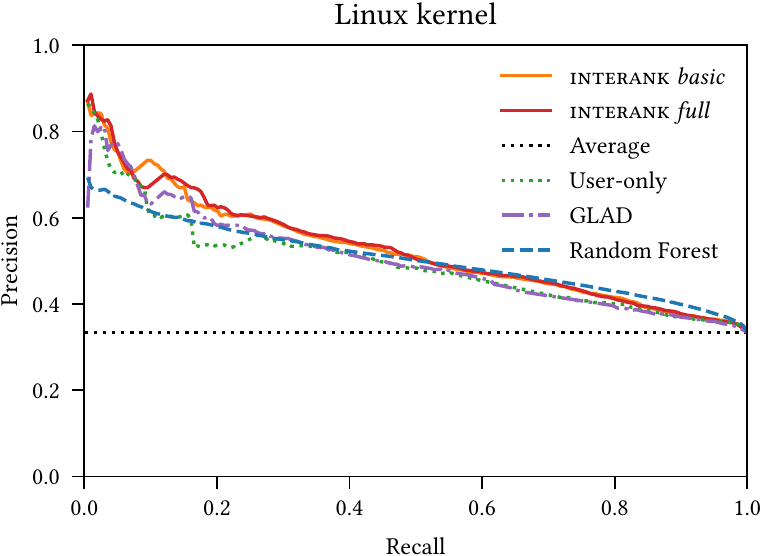}
  \caption{Precision-recall curves on the bad edit classification task for the Linux kernel. \textsc{interank} (solid orange and red) outperforms the user-only baseline (dotted green), the random forest classifier (dashed blue), and GLAD (dash-dotted purple).}
  \label{fig:linux_results}
\end{figure}

We show the precision-recall curves in Figure \ref{fig:linux_results}.
Both \interank{full} and \interank{basic} perform better than the four baselines.
Notably, they outperform the random forest in the high-precision regime, even though the random forest uses content-based features about developers, subsystems and patches.
In the high-recall regime, the random forest attains a marginally better precision.
The \emph{user-only} and GLAD baselines perform worse than all non-trivial models.

\subsection{Interpretation of Model Parameters}

We show in Table \ref{tab:linux_subsystems} the top-five and bottom-five subsystems according to  difficulties $\{d_i\}$ learned by \interank{full}.
We note that even though patches submitted to difficult subsystems have in general low acceptance rate, \interank{} enables a finer ranking by taking into account \emph{who} is contributing to the subsystems.
This effect is even more noticeable with the five subsystems with smallest difficulty value.

The subsystems $i$ with largest $d_i$ are \emph{core} components, whose integrity is crucial to the system.
For instance, the \texttt{usr} subsystem, providing code for RAM-related instructions at booting time, has barely changed in the last seven years.
On the other hand, the subsystems $i$ with smallest $d_i$ are \textit{peripheral} components serving specific devices, such as digital signal processors or gaming consoles.
These components can arguably tolerate a higher rate of bugs, and hence they evolve more frequently.

\begin{table}
  \caption{Top-five and bottom-five subsystems according to their difficulty $d_i$.}
  \label{tab:linux_subsystems}
  \centering
  \begin{tabular}{llrrr}
    \toprule
    Difficulty & Subsystem                  & \% Acc. & \# Patch & \# Dev.\\
    \midrule
    +2.664     & \texttt{usr}               & 1.88 \%  & 796      & 70 \\
    +1.327     & \texttt{include}           & 7.79 \%  & 398      & 101 \\
    +1.038     & \texttt{lib}               & 15.99 \% & 5642     & 707 \\
    +1.013     & \texttt{drivers/clk}       & 34.34 \% & 495      & 81 \\
    +0.865     & \texttt{include/trace}     & 17.73 \% & 547      & 81 \\
    \midrule
    -1.194     & \texttt{drivers/addi-data} & 78.31 \% & 272      & 8 \\
    -1.080     & \texttt{net/tipc}          & 43.11 \% & 573      & 44 \\
    -0.993     & \texttt{drivers/ps3}       & 44.26 \% & 61       & 9 \\
    -0.936     & \texttt{net/nfc}           & 73.04 \% & 204      & 26 \\
    -0.796     & \texttt{arch/mn10300}      & 45.40 \% & 359      & 63 \\
    \bottomrule
  \end{tabular}
\end{table}

\citet{jiang2013will} establish that a high prior subsystem churn (i.e., high number of previous commits to a subsystem) leads to lower acceptance rate.
We approximate the number of commits to a subsystem as the number of patches submitted multiplied by the subsystem's acceptance rate.
The first quartile of subsystems according to their increasing difficulty, i.e., the least difficult subsystems, has an average churn of \num{687}.
The third quartile, i.e., the most difficult subsystems, has an average churn of \num{833}.
We verify hence that higher churn correlates with difficult subsystems.
This corroborates the results obtained by \citeauthor{jiang2013will}

As shown in Figure \ref{fig:linux_results}, if false negatives are not a priority, \interank{} will yield a substantially higher precision.
In other words, if the task at hand requires that the patches classified as accepted are actually the ones integrated in a future release, then \interank{} will yield more accurate results.
For instance, it would be efficient in supporting Linus Torvalds in the development of the Linux kernel by providing him with a restricted list of patches that are likely to be integrated in the next release of the Linux kernel.

\section{Conclusion}
\label{sec:conclusion}

In this paper, we have introduced \interank{}, a model of edit outcomes in peer-production systems.
Predictions generated by our model can be used to prioritize the work of project maintainers by identifying contributions that are of high or low quality.

Similarly to user reputation systems, \interank{} is simple, easy to interpret and applicable to a wide range of domains.
Whereas user reputation systems are usually not competitive with specialized edit quality predictors tailored to a particular peer-production system, \interank{} is able to bridge the gap between the two types of approaches, and it attains a predictive performance that is competitive with the state of the art---without access to content-based features.

We have demonstrated the performance of the model on two peer-production systems exhibiting different characteristics.
Beyond predictive performance, we can also use model parameters to gain insight into the system.
On Wikipedia, we have shown that the model identifies controversial articles, and that latent dimensions learned by our model display interesting patterns related to cultural distinctions between articles.
On the Linux kernel, we have shown that inspecting model parameters enables to identify core subsystems (large difficulty parameters) from peripheral components (small difficulty parameters).

\paragraph{Future Work}
In the future, we would like to investigate the idea of using the latent embeddings learned by our model in order to recommend items to edit.
Ideally, we could match items that need to be edited with users that are most suitable for the task.
For Wikipedia, an ad-hoc method called ``SuggestBot'' was proposed by \citet{cosley2007suggestbot}.
We believe it would be valuable to propose a method that is applicable to peer-production systems in general.

\paragraph{Acknowledgments}
We are grateful to Yujuan Jiang for providing the Linux data and to Aaron Halfaker for helping us understand ORES.
We thank Patrick Thiran, Brunella Spinelli, Vincent Etter, Ksenia Konyushkova, Holly Cogliati-Bauereis and the anonymous reviewers for careful proofreading and constructive feedback.

\bibliography{wikirank} 

\end{document}